\documentclass[twocolumn,twocolumn]{IEEEtran}
\usepackage[T1]{fontenc}
\usepackage{color}
\usepackage{float}
\usepackage{mathrsfs}
\usepackage{amsmath}
\usepackage{amssymb}
\usepackage{graphicx}

\makeatletter

\floatstyle{ruled}
\newfloat{algorithm}{tbp}{loa}
\providecommand{\algorithmname}{Algorithm}
\floatname{algorithm}{\protect\algorithmname}

\usepackage{emptypage}

\usepackage[caption=false,font=footnotesize]{subfig}
\usepackage{algorithm}
\usepackage{algorithmic}

\usepackage{multirow} 
\usepackage{amsmath} 
\usepackage{xcolor}

\allowdisplaybreaks[4]

\ifCLASSOPTIONcompsoc
\usepackage[caption=false,font=normalsize,labelfont=sf,textfont=sf]{subfig}
\else
\usepackage[caption=false,font=footnotesize]{subfig}
\fi

\usepackage{cite}
\usepackage{bm}
\usepackage{algorithmic}
\usepackage{algorithm}
\usepackage{graphicx}
\IEEEoverridecommandlockouts

\usepackage{lettrine}

\usepackage{geometry}
\@ifundefined{showcaptionsetup}{}{%
 \PassOptionsToPackage{caption=false}{subfig}}
\usepackage{subfig}
\makeatother
\begin{document}
\title{Delay Optimization in Remote ID-Based UAV Communication via BLE and Wi-Fi Switching}
\author{\IEEEauthorblockN{Yian Zhu$^{\ast}$, Ziye Jia$^{\ast}$, Lei Zhang$^{\ast}$, Yao Wu$^{\ast}$, Qiuming Zhu$^{\ast}$, and Qihui Wu$^{\ast}$  \\
 }\IEEEauthorblockA{$^{\ast}$The Key Laboratory of Dynamic Cognitive System of Electromagnetic Spectrum Space, Ministry of Industry and Information Technology, Nanjing University of Aeronautics and Astronautics, 211106, China}\\ E-mail: \{zhuyian, jiaziye, Zhang\_lei, wu\_yao, zhuqiuming, wuqihui\}@nuaa.edu.cn
\thanks{This work was supported in part by National Natural Science Foundation of China under Grant 62301251 and 62231015, in part by the Aeronautical Science Foundation of China 2023Z071052007, and  in part by the Young Elite Scientists Sponsorship Program by CAST 2023QNRC001.\\ (\textit{Corresponding author: Ziye Jia})}}

\maketitle
\thispagestyle{empty}
\begin{abstract}
The remote identification (Remote ID) broadcast capability allows unmanned aerial vehicles (UAVs) to exchange messages, which is a pivotal technology for inter-UAV communications. Although this capability enhances the operational visibility, low delay in Remote ID-based communications is critical for ensuring the efficiency and timeliness of multi-UAV operations in dynamic environments. To address this challenge, we first establish delay models for Remote ID communications by considering packet reception and collisions across both BLE 4 and Wi-Fi protocols. Building upon these models, we formulate an optimization problem to minimize the long-term communication delay through adaptive protocol selection. 
Since the delay performance varies with the UAV density, we propose an adaptive BLE/Wi-Fi switching algorithm based on the multi-agent deep Q-network approach. 
Experimental results demonstrate that in dynamic-density scenarios, our strategy achieves 32.1\% and 37.7\% lower latency compared to static BLE 4 and Wi-Fi modes respectively.
 \end{abstract}
\vspace{-3mm}
\newcommand{\CLASSINPUTtoptextmargin}{0.8in}
\newcommand{\CLASSINPUTbottomtextmargin}{1in}
\section{Introduction}
\lettrine[lines=2]
T{he} rapid development of unmanned aerial vehicles (UAVs) has prompted governments and institutions to establish secure regulations for UAV operations \cite{DONG2024}, \cite{10599389}, \cite{10616106}, \cite{11006480}. To address the need, the federal aviation administration (FAA) has explored various mechanisms for real-time UAV tracking, focusing on transmitting identity and status information. As part of these efforts, FAA introduced the remote identification (Remote ID) technology, which enables UAV tracking through cellular or radio-frequency (RF) communication links \cite{10311133}. 
However, due to the limited reliability of cellular infrastructure in many regions and the associated cost constraints, FAA has opted for low-power communication protocols as the standard for Remote ID. Specifically, the bluetooth low energy (BLE) and wireless fidelity (Wi-Fi) have been designated as the primary technologies for the Remote ID compliance \cite{astm13}.

The Remote ID system utilizes the onboard sensors of UAVs, such as global navigation satellite system (GNSS) and inertial measurement unit, to collect real-time data on latitude, longitude, altitude, speed, and system status. Then, the data is periodically broadcast in standardized message packets via BLE or Wi-Fi communication protocols. Currently, Remote ID has been mandatorily deployed through legislation in countries such as United States and European Union, establishing a traceable supervision system of operating UAVs \cite{9868084}.
Notably, Remote ID is not only a regulatory tool for compliance monitoring but also a technical enabler for advanced UAV applications. The broadcast mechanism of Remote ID inherently supports decentralized airspace awareness through real-time inter-UAV message exchange. By receiving Remote ID messages broadcast by nearby UAVs, UAVs can autonomously build a decentralized UAV-to-UAV awareness network. This capability provides foundational communication supports for multi-UAV collaborative tasks without relying on additional infrastructures.
For instance, in \cite{9121767}, broadcasting the mission status and positional data among UAVs enables efficient task coordination and collision avoidance during multi-UAV object delivery and data collection operations. \cite{9594483} leverages real-time position exchange among UAVs to implement a distributed conflict warning system. These implementations highlight the potential of Remote ID as a standardized communication layer for advanced applications.

However, due to the highly mobile nature of UAVs, low latency is critical for applications such as conflict detection, collision avoidance, and task coordination \cite{wu2023latency}, \cite{9394785}. Excessive latency can lead to delayed decision-making, increased risk of collisions, and degraded performance in collaborative missions. Therefore, analyzing and optimizing the latency of communication methods is essential to ensure the reliable and efficient operation of UAV systems. Recent studies have explored latency optimization in related communication protocols. For instance, \cite{7458902} models the neighbor discovery process in BLE 4, considering factors such as signal collisions, discovery delay, and energy consumption. However, it neglects the critical aspect of dynamic mobility exhibited by UAV nodes in aerial communication scenarios. \cite{10366188} introduces a BLE frequency-hopping algorithm to mitigate Wi-Fi signal collisions, but it focuses primarily on resolving packet collisions rather than analyzing or optimizing the broadcast protocols of BLE and Wi-Fi. \cite{7986413} evaluates Wi-Fi performance in multi-UAV communications, focusing on packet loss, but it lacks a theoretical model for communication protocols and relies solely on the experimental validation. \cite{10322686} proposes a hybrid BLE/LTE/Wi-Fi/LoRa switching scheme for energy and delay optimization in UAV networks, but it does not provide a detailed model of the protocol mechanism and overlooks the impact of interference and transmission frequency on latency.

Building on the insights from the aforementioned studies, we investigate the communication delay optimization problem in UAV networks based on Remote ID. Specifically, we analyze the delay characteristics of two Remote ID communication protocols, BLE 4 and Wi-Fi, considering detailed transmission mechanisms and interference effects. The objective is to minimize the long-term system average message delay for UAV Remote ID. To achieve this, we design the multi-agent deep Q-network-based BLE/Wi-Fi switching algorithm (MADQN-BWSA) to dynamically optimize the protocol selection and configuration for UAVs. Simulation results show that the proposed algorithm outperforms baseline methods in terms of delay performance in dynamic UAV networks.

\begin{figure}[t]
  \centering
  \includegraphics[scale=0.37]{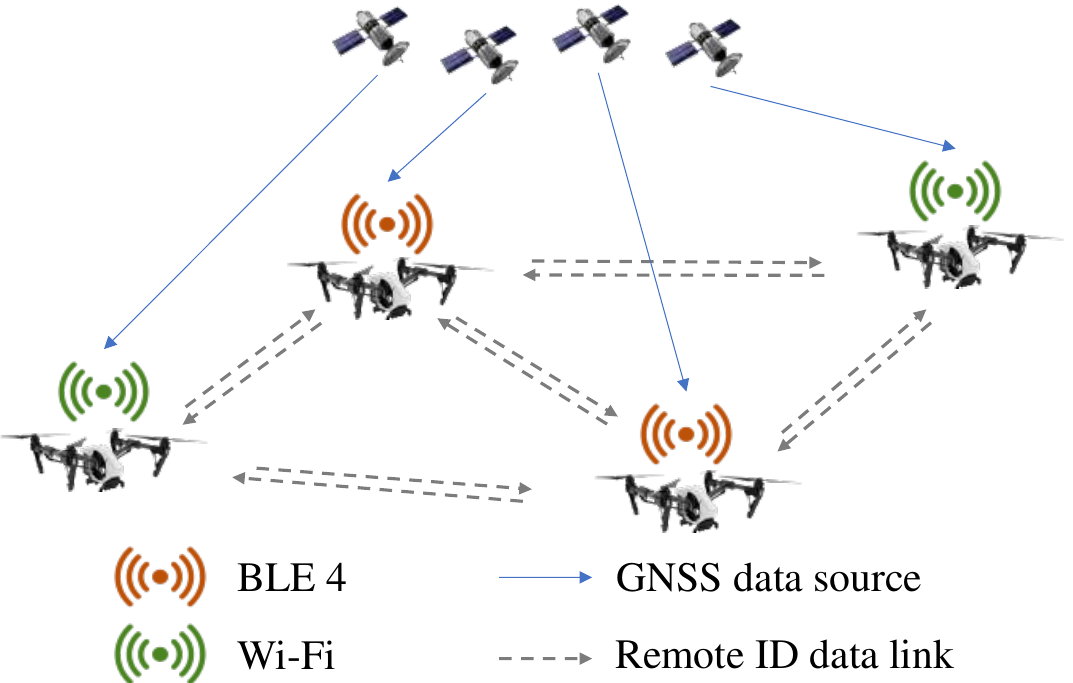}\caption{\label{fig:scenario}Remote ID-based UAV communication networks.}\vspace{-3mm}
  \end{figure}
\section{System Model and Problem Formulation\label{sec:sec2}}
As depicted in Fig. \ref{fig:scenario}, GNSS serves as the data source for Remote ID, providing precise status information for UAVs. UAVs utilize Remote ID to broadcast messages to others, employing either BLE 4 or Wi-Fi as the underlying communication protocol. To optimize the communication performance of such system, this section analyzes the average delays of Remote ID and formulates the optimization problem.
\subsection{Packet Reception Model}\label{sec:latencymodel}
We evaluate the packet reception delay for BLE 4 and Wi-Fi protocols using a discrete-time slot model, where a time period is divided into discrete slots with length $\Delta$. 
\subsubsection[short]{BLE 4}\label{sec:BLE4}
\begin{figure}[t]
  \centering
  \begin{minipage}{\linewidth}
      \centering
      \subfloat[BLE 4 operational mode.]{
          \includegraphics[width=0.65\linewidth]{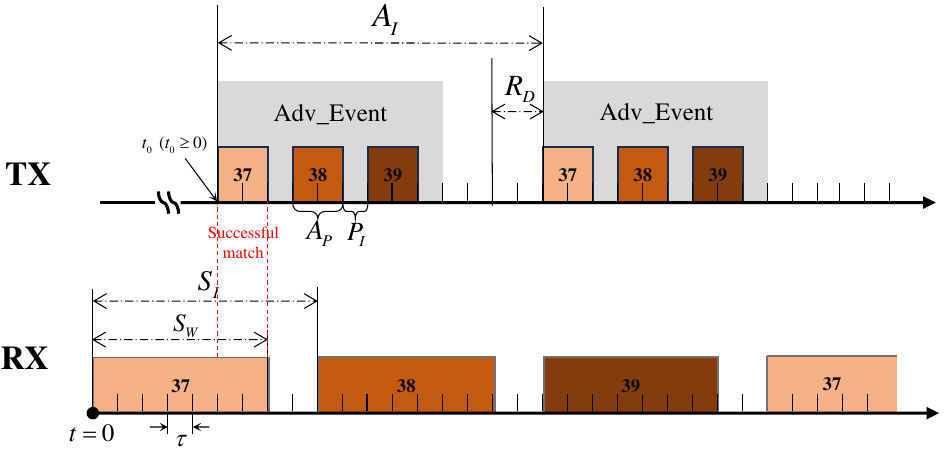}
          \label{fig:legacyoperation}
      }\vspace{-2mm}
      \vfill
      \subfloat[Wi-Fi operational mode.]{
      \includegraphics[width=0.65\linewidth]{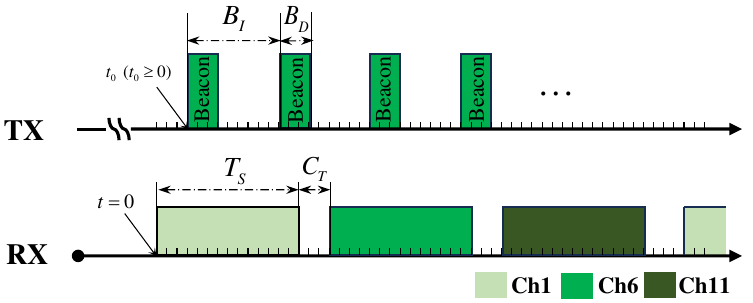}
     
      \label{fig:wifioperation}
      }
    
\caption{Operational mode of communication protocols for BLE 4 and Wi-Fi.}
      \label{fig:operation model}
  \end{minipage}
\vspace{-3mm}
\end{figure}

As shown in Fig. \ref{fig:operation model}(a), the BLE 4 broadcast process involves a transmitter (TX) generating a broadcast event (Adv\_Event) consisting of three identical protocol data unit (PDU) packets, transmitted sequentially on channels 37, 38, and 39. Each PDU packet carries the actual Remote ID data and has a duration of $A_P$, with an interval $P_I$ between packets. Consecutive broadcast events are separated by $A_I$, which includes a pseudo-random delay $R_D$. The receiver (RX) scans periodically on the same channels, with each scan window lasting $S_W$ and repeating every $S_I$.

To determine when the broadcast event and the scanning window align, the Chinese remainder theorem (CRT) \cite{CRT} is used to solve the congruence equations and identify matching time slots. Specifically, the CRT provides the solution for cases where two periodic events, such as BLE 4 transmission and scanning on the corresponding channels, align over a common period. The CRT can be expressed as

\begin{equation}\label{eq:CRT}
  \begin{aligned}
    &\mathcal{T}\left(t_1, t_2, \mathcal{S}_1, \mathcal{S}_2\right)=\left\{\Delta_o+k \mathcal{S}_1 \mathcal{S}_2 \mid k \in \mathbb{N}\right\},\\ 
    &\Delta_o=\left(t_1 \mathcal{S}_2\left[\mathcal{S}_2^{-1}\right]_{\mathcal{S}_1}+t_2 \mathcal{S}_1\left[\mathcal{S}_1^{-1}\right]_{\mathcal{S}_2}\right) \bmod \left(\mathcal{S}_1 \mathcal{S}_2\right),
  \end{aligned}
\end{equation}
where $t_1$ and $t_2$ are the start times of the two periodic events, $\mathcal{S}_1$ and $\mathcal{S}_2$ are the event periods, and $[\mathcal{S}_1^{-1}]_{\mathcal{S}_2}$ is the modular multiplicative inverse of $\mathcal{S}_1$ modeulo $\mathcal{S}_2$.

We define $\psi_{\text{BLE 4}}$ as the number of Remote ID messages transmitted per GNSS update cycle, representing the message transmission rate. Assuming the initial GNSS update time is $t_0$ ($t_0 > 0$), the set of successful matching time slots for the three BLE 4 broadcast channels $c \in \{37, 38, 39\}$ during a single GNSS update cycle is 
\begin{equation}\label{eq:leg3}
  \begin{gathered}
  \mathcal{M}_{\text{BLE 4}}^{c}\!(t_0)\! =\! \left\{\! t \!\mid\! t \!\in \!\mathcal{T}\!\left( \!t_0 \!+\! (c-37)(A_P \!+\! P_I),\! i\!,\! \hat{A}_I\!, \!3 S_I\! \right), \!\right. \\
  (c\!-\!37) \!S_I\! \leq \!i\! \leq\! (c-37)\! S_I \!+\! S_W \!- \!A_P, \left. \!t_0 \!\leq t\! \leq\! t_0 \!+\! T_{\text{GNSS}} \right\}. \\
  \end{gathered}
\end{equation}
Here, $A_I = \frac{T_{\text{GNSS}}}{\psi_{\text{BLE 4}}}$, where $T_{\text{GNSS}}$ is the GNSS update period. To apply the CRT, $A_I$ must be relatively prime to $3 S_I$. Therefore, $A_I$ is approximated by $\hat{A}_I$, the closest integer that satisfies this constraint.
The complete set of matching time slots for all three channels is
\begin{equation}\label{eq:BLE4matchset}
  \mathcal{N}_{\text{BLE 4}}(t_0) = \left\{ \mathcal{M}_{\text{BLE 4}}^{37}(t_0), \mathcal{M}_{\text{BLE 4}}^{38}(t_0), \mathcal{M}_{\text{BLE 4}}^{39}(t_0) \right\}.
\end{equation}
Finally, the reception delay for each successfully received PDU packet is
\begin{equation}\label{eq:BLE4delay} D_{\text{BLE 4}}(\delta) = \delta - t_0 + A_P, \quad \delta \in \mathcal{N}_{\text{BLE 4}}(t_0). \end{equation}
\subsubsection{Wi-Fi}\label{sec:wifi1}
The Wi-Fi broadcasting mechanism is depicted in Fig. \ref{fig:operation model}(b). The UAV selects a fixed channel for broadcasting, designated as channel 6 in this case. The beacon packet, which carries the Remote ID data, has a duration of $B_D$, and the interval between successive beacons is $B_I = \frac{T_{\text{GNSS}}}{\psi_{\text{Wi-Fi}}}$, where $\psi_{\text{Wi-Fi}}$ is the message transmission rate of Wi-Fi.

The RX employs passive scanning, listening on channels 1, 6, and 11 for a duration of $T_S$ to capture the beacon packet before switching to the next channel. The time to switch between channels is denoted as $C_T$.  Using (\ref{eq:CRT}), the set of matching time slots for beacon packets on channel $c$ $(c \in\{1,6,11\})$ is
  \begin{equation}\label{eq:bcndelay}
    \begin{gathered}
    \mathcal{M}^c_{\text{Wi-Fi}}(t_0) = \{ t \mid t \in \mathcal{T}(t_0, i, \hat{B}_I, 3(T_S + C_T)), i \in I_c, \\
    t_0 \leq t \leq t_0 + T_{\text{GNSS}} \},
    \end{gathered}
  \end{equation}
where $\hat{B}_I$ is an approximation of $B_I$ that is relatively prime to $3(T_S + C_T)$. The range of time slot values $I_c$ for channel $c$ is defined as
\begin{equation}
  I_c= \begin{cases}0 \leq i \leq T_S-B_D, & \text { if } c=1, \\ T_S+C_T \leq i \leq 2 T_S+C_T-B_D, & \text { if } c=6, \\ 2\left(T_S+C_T\right) \leq i \leq 3 T_S+2 C_T-B_D, & \text { if } c=11.\end{cases}
  \end{equation}
Thus, the reception delay for the Wi-Fi beacon packet on channel $c$ is
  \begin{equation}\label{eq:bcn}
    D_{\text{Wi-Fi}}(\delta) = \delta - t_0 + B_D, \quad \delta \in \mathcal{M}_{\text{Wi-Fi}}^c(t_0).
  \end{equation}
\subsection{Packet Collision Model\label{sec:packetcollision-Model}}
\subsubsection{BLE 4}\label{sec:blesticti}
Let $\mathcal{U}=\left\{u_1, u_2, \ldots, u_M\right\}$ represent the set of $M$ UAVs. The set of UAVs from which UAV $u_i \in \mathcal{U}$ can receive Remote ID messages via BLE 4 is defined as
\begin{equation}
  \mathcal{R}_\text{BLE 4}^{u_i}\!\!=\!\!\left\{u_j \!\mid\! u_j \in \mathcal{U}, i_{u_j}^\text{BLE 4} \!\!\cdot\!\!\left(P_{u_j}^{\text{BLE 4}, \text{tx}}\!-\!\mathrm{PL}_{u_j, u_i}\right) \!\geq\! \Theta_\text{BLE 4}\right\},
  \end{equation}
  where $i_{u_j}^{\text{BLE 4}}$ indicates whether UAV $u_j$ uses BLE 4, $P_{u_j}^{\text{BLE 4,tx}}$ is its transmit power, and $\mathrm{PL}_{u_j, u_i}$ is the path loss between UAV $u_j$ and $u_i$. $\psi_{\text {BLE 4}}^{u_j}$ represents the Remote ID message transmission rate for UAV ${u_j}$ when using BLE 4.

  The probability that UAV $u_i$ successfully receives a single PDU packet from UAV $u_j \in \mathcal{R}_{\text {BLE 4}}^{u_i}$ without self-technology interference (STI) is 
  \begin{equation}
    P_{\bar{c}, u_j, u_i}^{\text{BLE 4}, \text{STI}} = \prod_{\substack{u_k \in \mathcal{R}_{\text{BLE 4}}^{u_i} \\ u_k  \neq u_j}} \left( 1 - \frac{\psi_{\text{BLE 4}}^{u_k} \cdot 2 A_P}{T_{\text{GNSS}}} \right).
  \end{equation}
\subsubsection{Wi-Fi}\label{sec:wifiSTICTI}
To define the STI for Wi-Fi, we use $c$ to differentiate UAVs within the vicinity of $u_i$ that use different Wi-Fi channels to send beacons, where $u_j \in \mathcal{R}_{\text{\scriptsize{Wi-Fi}}}^{u_i, c}$ and $c \in\{1,6,11\}$. The set $\mathcal{R}_{\text{\scriptsize{Wi-Fi}}}^{u_i, c}$ is analogous to $ \mathcal{R}_\text{BLE 4}^{u_i}$, representing the set of UAVs within the range of $u_i$ transmitting Wi-Fi beacons on channel $c$.  The non-collision probability that UAV $u_i$ successfully receives a Wi-Fi beacon packet sent by UAV $u_j \in \mathcal{R}_{\text{\scriptsize{Wi-Fi}}}^{u_{i, c}}$ under Wi-Fi interference is 
\begin{equation}
  P_{\bar{c}, u_j, u_i}^{\text{Wi-Fi}, \text{STI}} = \prod_{\substack{{u_k} \in \mathcal{R}_{\text{Wi-Fi}}^{u_i, c} \\ {u_k} \neq u_j}} \left( 1 - \frac{\psi_{\text{Wi-Fi}}^{u_k} \cdot 2 B_D}{T_{\text{GNSS}}} \right).  
\end{equation}

\subsection{Average Transmission Delay Model\label{sec:sec4}}
In this section, we model the average transmission delay for both BLE 4 and Wi-Fi protocols.
\subsubsection{BLE 4}
For UAV $u_j \in \mathcal{U}$, the set of neighboring UAVs capable of successfully receiving Remote ID messages transmitted via BLE 4 is defined as
\begin{equation}
  \mathcal{S}_{\text{BLE 4}}^{u_j} \!\!= \!\left\{ u_i \!\mid\! u_i \in \mathcal{U}, i_{u_j}^{\text{BLE 4}} \left( P_{u_j}^{\text{BLE 4,tx}} \!-\! \!\mathrm{PL}_{u_j, u_i} \right) \!\geq \!\!\Theta_{\text{BLE 4}} \right\}.
\end{equation} 

When $u_j$ broadcasts via BLE 4, the set of successfully received PDU packets within a GNSS update cycle starting at $t_0$ is denoted as $\mathcal{N}_{\text {BLE } 4}^{u_j}\left(t_0\right)$ in (\ref{eq:BLE4matchset}). Let $\delta_{u_j, n}^{\text {BLE } 4}$ represent the $n$-th earliest packet reception time among all successfully received packets, i.e., $\delta_{u_j,n}^{\text {BLE 4}} = \min \{\mathcal{N}_{\text {BLE 4}}^{u_j}(t_0)\}_n$. Considering the packet reception delay and the collision-free probability of BLE 4 packets, the BLE 4 transmission delay from $u_j \in \mathcal{U}$ to a neighboring UAV $u_i \in \mathcal{S}_{\text {BLE } 4}^{u_j}$ at time $t_0$ can be expressed as
\begin{equation}\label{eq:Legdelay}
            \begin{aligned}
              d_{\text{BLE 4}}^{u_j, u_i}\left(t_0\right) &= \sum_{k=0}^{\infty} \sum_{n=1}^{\left| N_{\text{BLE 4}}^{u_j}\left(t_0\right) \right|} 
              \left( 1 - P_{\bar{c}, u_j, u_i}^{\text{BLE 4}, \text{STI}}  \right)^{n-1 + \left| N_{\text{BLE 4}}^{u_j}\left(t_0\right) \right| \times k} \\
              & \quad \times P_{\bar{c}, u_j, u_i}^{\text{BLE 4}, \text{STI}}  \times \left( D\left( \delta_{u_j,n}^{\text{BLE 4}} \right) + k \times T_{\text{GNSS}} \right),
            \end{aligned}
\end{equation}  
where $|\cdot|$ represents the cardinality of the corresponding set.

Considering the stochastic nature of packet transmission timing, the average delay for Remote ID messages sent by $u_j$ to $u_i \in \mathcal{S}_{\text {BLE 4}}^{u_j}$ is
      \begin{equation}
        \bar{d}_{\text{BLE 4}}^{u_j,u_i}=\frac{1}{3 S_I} \sum_{t_0=0}^{3 S_I-1} d_{\text{BLE 4}}^{u_j,u_i}\left(t_0\right).
        \end{equation}  \vspace{-2mm}
\subsubsection{Wi-Fi}
Similar to the definition of $\mathcal{S}_{\text{BLE 4}}^{u_j}$, $\mathcal{S}_{\text{Wi-Fi}}^{u_j}$ represents the set of neighboring UAVs that can successfully receive the Remote ID message transmitted via Wi-Fi. Using (\ref{eq:bcndelay}), the set of matching time slots, denoted as $\mathcal{M}_{\text{Wi-Fi}}^{c, u_j}\left(t_0\right)$, is obtained, where the beacon packet transmitted by $u_j$ is successfully received. Let $\delta_{u_j, n}^{\text{Wi-Fi}}$ represent the $n$-th smallest time slot in $\mathcal{M}_{\text{Wi-Fi}}^{c, u_j}\left(t_0\right)$.
By combining the packet reception delay with the packet collision probability model, the Wi-Fi transmission delay from UAV $u_j$ at time $t_0$ to $\operatorname{UAV} u_i \in \mathcal{S}_{\text{Wi-Fi}}^{u_j}$ is 
  \begin{equation}\label{eq:wifidelay}
    \begin{aligned}
    & d_{\text{Wi-Fi}}^{u_j, u_i}\left(t_0\right) = \sum_{k=0}^{\infty} \sum_{n=1}^{\left| \mathcal{M}_{\text{Wi-Fi}}^{c, u_j}\left(t_0\right) \right|} \left( 1 -  P_{\bar{c}, u_j, u_i}^{\text{Wi-Fi}, \text{STI}}  \right)^{n - 1 + \left| \mathcal{M}_{\text{Wi-Fi}}^{c, u_j}\left(t_0\right) \right| \times k} \\
    & \times  P_{\bar{c}, u_j, u_i}^{\text{Wi-Fi}, \text{STI}}  \times \left( D\left( \delta_{u_j, n}^{\text{Wi-Fi}} \right) + k \times T_{\text{GNSS}} \right).
    \end{aligned}
  \end{equation}

  Considering the stochastic nature of packet transmission timing, the average Remote ID message delay for $u_i \in \mathcal{S}_{\text{Wi-Fi}}^{u_j}$ from by $u_j$ using Wi-Fi is
  \begin{equation}
    \bar{d}_{\text{Wi-Fi}}^{u_j,u_i} = \frac{1}{3\left(T_S + C_T\right)} \sum_{t_0 = 0}^{3\left(T_S + C_T\right) - 1} d_{\text{Wi-Fi}}^{u_j,u_i}\left(t_0\right).
  \end{equation}

\subsection{Problem Formulation}
The objective is to minimize the long-term average message transmission delay for each UAV by jointly optimizing the transmission protocol selection $\mathcal{I}(t)=\{i_{u_j}^\epsilon(t)\mid\forall u_j \in \mathcal{U}, \epsilon \in \mathcal{E}\}$ and the message transmission rate $\Psi(t)=\{\psi_\epsilon^{u_j}(t)\mid\forall u_j \in \mathcal{U},\epsilon \in \mathcal{E}\}$, where $\mathcal{E}=$\{BLE 4, Wi-Fi\}. The optimization problem is formulated as
\begin{subequations}\label{eq:ctr_shale}
  \begin{align}
    \text{P0:}\min_{\mathcal{I}(t), \Psi(t)} & \frac{1}{T_{\max}} \sum_{t \in \mathcal{T}} \sum_{u_j \in \mathcal{U}} \frac{1}{|S_{u_j}|} \sum_{u_i \in S_{u_j}}\sum_{\epsilon \in \mathcal{E}}\left(\bar{d}_\epsilon^{u_j, u_i}(t) \cdot i_{u_j}^\epsilon\right) \tag{\ref{eq:ctr_shale}} \\
  \text{s.t.} \quad & \sum_{\epsilon \in \mathcal{E}} i_{u_j}^{\epsilon}(t) = 1, \quad \forall u_j \in \mathcal{U}, \\
  & \psi_{\epsilon}^{u_j}(t) \leq \psi_{\epsilon, \max}, \quad \forall u_j \in \mathcal{U}, \epsilon \in \mathcal{E}, \\
  & i_{u_j}^{\epsilon}(t) \in \{0, 1\}, \quad \forall u_j \in \mathcal{U}, \epsilon \in \mathcal{E},\\
  & \psi_{\epsilon}^{u_j}(t) \in \mathbb{Z}^+, \quad \forall u_j \in \mathcal{U}, \epsilon \in \mathcal{E}.
  \end{align}
  \end{subequations}
  Wherein, $T_{\max }$ denotes the total observation duration, and $\mathcal{T}$ represents the set of all time points, $S_{u_j}=\left\{\mathcal{S}_{\text{BLE 4}}^{u_j}  \cup \mathcal{S}_{\text{Wi-Fi}}^{u_j}\right\}$. Constraint (\ref{eq:ctr_shale}a) ensures that each UAV $u_j$ selects exactly one communication protocol from the set $\mathcal{E}$ at $t$. Constraint (\ref{eq:ctr_shale}b) limits the maximum message transmission rate $\psi_{\epsilon, \max}$ for each communication technology $\epsilon$. Constraint (\ref{eq:ctr_shale}c) enforces a binary decision for the communication protocol selection. Constraint (\ref{eq:ctr_shale}d) ensures that the message transmission rate $\psi_\epsilon^{u_j}(t)$ is a positive integer. 
  
P0 involves system-wide decision-making for UAVs in dynamic environments, formulated as a mixed-integer nonlinear programming problem. Traditional optimization methods struggle to handle its combinatorial complexity and real-time requirements.
\vspace{-3mm}
\section{Algorithm Design}\label{sec:algo}
 To address P0, it is transformed as a Markov decision process (MDP), and then a DRL-based MADQN-BWSA is designed.
\subsection{MDP Design}\label{sec:design of MDP}
\subsubsection{State for each UAV}
At time $t$, each UAV $u_j \in \mathcal{U}$ obtains a local observation $o_{u_j}(t)$ from the environment, which includes information from nearby UAVs. In detail,
\begin{equation} 
  o_{u_j}(t) \triangleq \left( \mathbf{I}_{u_j}^s(t), \mathbf{\Psi}_{u_j}^s(t), \mathbf{D}^{\mathcal{R}^{u_j}}(t), \mathbf{I}^{\mathcal{R}^{u_j}}(t), \mathbf{\Psi}^{\mathcal{R}^{u_j}}(t) \right), 
\end{equation}
where $\mathbf{I}_{u_j}^s(t)=\left\{i_{u_j}^\epsilon(t) \mid \epsilon \in \mathcal{E}\right\}$ is the communication protocols selected by UAV $u_j$ at time $t$.
$\boldsymbol{\Psi}_{u_j}^s(t)=\left\{\psi_\epsilon^{u_j}(t) \mid \epsilon \in \mathcal{E}\right\}$ denotes the message transmission rates corresponding to the selected protocols.
$\mathbf{D}^{\mathcal{R}^{u_j}}(t)=\left\{d_{u_k, u_j}(t) \mid u_k \in \bigcup_{\epsilon \in \mathcal{E}} \mathcal{R}_\epsilon^{u_j}\right\}$ is the distances between UAV $u_j$ and UAV $u_k$.
  $\mathbf{I}^{\mathcal{R}^{u_j}}(t)=\left\{i_{u_k}^\epsilon(t) \mid \epsilon \in \mathcal{E}, u_k \in \bigcup_{\epsilon \in \mathcal{E}} \mathcal{R}_\epsilon^{u_j}\right\}$ is the set of protocols used by UAV $u_k$ transmitting messages to UAV $u_j$.
 $\boldsymbol{\Psi}^{\mathcal{R}^{u_j}}(t)=\left\{\psi_\epsilon^{u_k}(t) \mid \epsilon \in \mathcal{E}, u_k \in \bigcup_{\epsilon \in \mathcal{E}} \mathcal{R}_\epsilon^{u_j}\right\}$ indicates the message transmission rates of transmitting UAV $u_k$.
 
\subsubsection{Action for each UAV}
Based on the observation $o_j(t)$, the action of UAV $u_j \in \mathcal{U}$ at time $t$ is defined as
\begin{equation}
  a_{u_j}(t) \triangleq\left(\mathbf{I}_{u_j}^a(t), \mathbf{\Psi}_{u_j}^a(t)\right),
  \end{equation}
where $\mathbf{I}_{u_j}^a(t)=\left\{i_{u_j}^{\epsilon, a}(t) \mid \epsilon \in \mathcal{E}\right\}$ denotes the selected communication protocols. $\boldsymbol{\Psi}_{u_j}^a(t)=\left\{\psi_\epsilon^{u_j, a}(t) \mid \epsilon \in \mathcal{E}\right\}$ represents the corresponding message transmission rates.
\subsubsection{Reward function of each UAV}
The reward function is designed to optimize Remote ID message transmission delays by considering both local transmission performance and global delay, which is defined as
\begin{equation}
r_{u_j}(t) = \alpha r_{u_j}^{\text{local}}(t) + \beta r_{u_j}^{\text{global}}(t),
\end{equation}
where $r_{u_j}^{\text{local}}(t)= -\frac{1}{\left|S_{u_j}\right|} \sum_{u_i \in S_{u_j}} \bar{d}^{u_j, u_i}(t)$ motivates UAV $u_j$ to minimize the message delays to its neighboring UAVs, enhancing the local communication efficiency, $r_{u_j}^{\text{global}}(t) =\frac{1}{\left|S_{u_j}\right|} \sum_{u_i \in S_{u_j}}  (\bar{d}_{\text{global}}(t) -\bar{d}^{u_j, u_i}(t)) $ prevents UAVs from making suboptimal decisions due to limited local observations, ensuring  the alignment with the global transmission average. Here, the global average transmission delay $\bar{d}_{\text{global}}(t) = \frac{1}{M} \sum_{u_j \in \mathcal{U}} \frac{1}{\left|S_{u_j}\right|} \sum_{u_i \in S_{u_j}} \bar{d}^{u_j, u_i}(t)$. $\alpha$ and $\beta$ are weighting factors that balance the impact of local and global delay.

\subsection{Training Algorithm}
In dynamic environments with communication constraints and incomplete global information, UAVs must rely on local observations for distributed decision-making. To address this, we adopt the MADQN algorithm, which supports the distributed training and execution.

Each agent in MADQN maintains two Q-networks: the primary Q-network $Q_{\theta_j}\left(o_{u_j}, a_{u_j}\right)$ and the target Q-network $Q_{\theta_j^{\prime}}\left(o_{u_j}, a_{u_j}\right)$, where $o_{u_j}$ represents the local observation of UAV $u_j$ and $a_{u_j}$ represents the action taken by $u_j$. The target Q-network stabilizes training by periodically updating its parameters $\theta_j^{\prime}$ to match those of the primary network $\theta_j$.

The MADQN-BWSA trains UAVs through distributed decision-making across multiple episodes. At each time step, each UAV selects actions using an epsilon-greedy strategy, balancing the exploration and exploitation. UAVs update their Q-networks by applying temporal difference learning, minimizing the loss function, and periodically updating the target Q-network via soft updates. Experiences are stored in a replay buffer, and Q-network parameters are updated after sufficient data has been accumulated. This process continues until the maximum number of episodes is reached, allowing UAVs to optimize transmission delays based on local observations. The MADQN-BWSA process is summarized in Algorithm \ref{alg:madqn_atmc}.
\begin{algorithm}[t]
  \caption{MADQN-BWSA}
  \begin{algorithmic}[1]
    \label{alg:madqn_atmc}
    \STATE \textbf{Input:} Local observations $o_{u_j}(t)$, maximum time steps  $T_{\text{max}}$, maximum number of episodes $E_{\text{max}}$, exploration rates $e_{\text{init}}$ and $e_{\text{final}}$, batch size $\mathcal{B}$, and replay buffer $D$ with capacity $D_{\text{max}}$.
    \STATE \textbf{Output:} Remote ID communication configurations for each UAV.
    \STATE \textbf{Initialize:} Q-networks $Q_{\theta_j}$, target Q-networks $Q_{\theta_j^{\prime}}$ for each UAV.
    \FOR{episode = 1 to max-episode}
        \STATE $e = \max(e_{\text{final}}, e_{\text{init}} - \frac{\text{episode} (e_{\text{init}} - e_{\text{final}})}{E_{\text{max}}})$.
        \FOR{t = 1 to $T_{\text{max}}$}
            \FOR{$j=1$ to $M$}
                \STATE Observe $o_{u_j}(t)$.
                \IF{$r < e$}
                    \STATE Select $a_{u_j}(t)$ as \text{random action}.
                \ELSE
                    \STATE Select $a_{u_j}(t)$ as $\arg\max_{a_{u_j}} Q_{\theta_j}(o_{u_j}(t), a_{u_j})$.
                \ENDIF
                \STATE Execute $a_{u_j}(t)$, receive $r_{u_j}(t)$, observe $o_{u_j}(t+1)$\!.
                \STATE Store $\langle o_{u_j}(t), a_{u_j}(t), r_{u_j}(t), o_{u_j}(t+1) \rangle$ in $D$.
            \ENDFOR
            \IF{$|D| \geq D_{\text{max}}$}
                \FOR{$j=1$ to $M$}
                    \STATE Sample $\mathcal{B}$ from $D$.
                    \STATE Update the Q-network: \\ $\mathcal{L}(\theta_j) = \frac{1}{|\mathcal{B}|} \sum_i \left[ y_j^i(t) - Q_{\theta_j}(o_{u_j}^i(t), a_{u_j}^i(t)) \right]^2$.
                    \STATE Update target Q-network: $\theta_j^{\prime} \leftarrow \tau \theta_j + (1 - \tau) \theta_j^{\prime}$.
                \ENDFOR
            \ENDIF
            \STATE Update UAV positions randomly.
        \ENDFOR
    \ENDFOR
  \end{algorithmic}
\end{algorithm}\vspace{-2mm}
\section{Simulation Results\label{sec:Simulation Results}}
\subsection{Analysis of Fixed Remote ID Mode}
We first evaluate the communication performance of fixed protocol Remote ID modes in a multi-UAV environment with 10 UAVs. The altitude of each UAV is restricted to 30 - 120m, and the horizontal flight space varies in different size. The flight speed of each UAV is limited to 20 m/s, and they follow random paths in this airspace. The sampling period of GNSS is set to 1s. Both BLE 4 and Wi-Fi have a data rate of 1Mbps and transmit power of 18dBm. BLE 4 has a receiver sensitivity of -85dBm,  while that of Wi-Fi is -105dBm. The path loss exponent is 2.1, and shadowing variance is 6dB, with a log-normal shadowing model applied. The value of $\varDelta$ is 0.125ms for both technologies. Additional parameters for BLE 4 include $A_P$=0.376ms, $P_I$=0.125 ms, $R_D$=5ms, $S_W$=2ms, and $S_I$=8ms, while Wi-Fi includes $B_D$=0.632ms, $T_S$=6ms, and $C_T$=1ms.

\begin{figure}[t]
  \centering
  \subfloat[]{\includegraphics[width=0.24\textwidth]{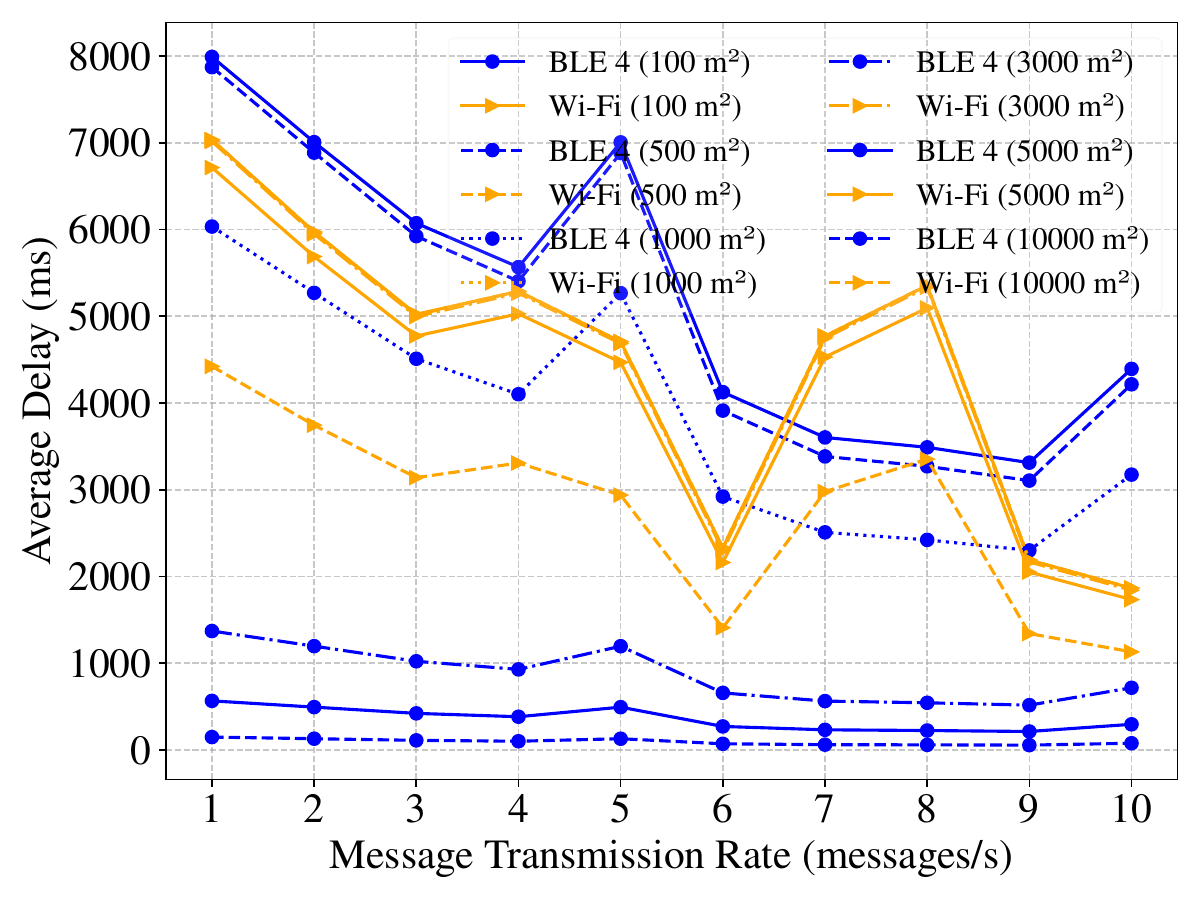}}
  \subfloat[]{\includegraphics[width=0.24\textwidth]{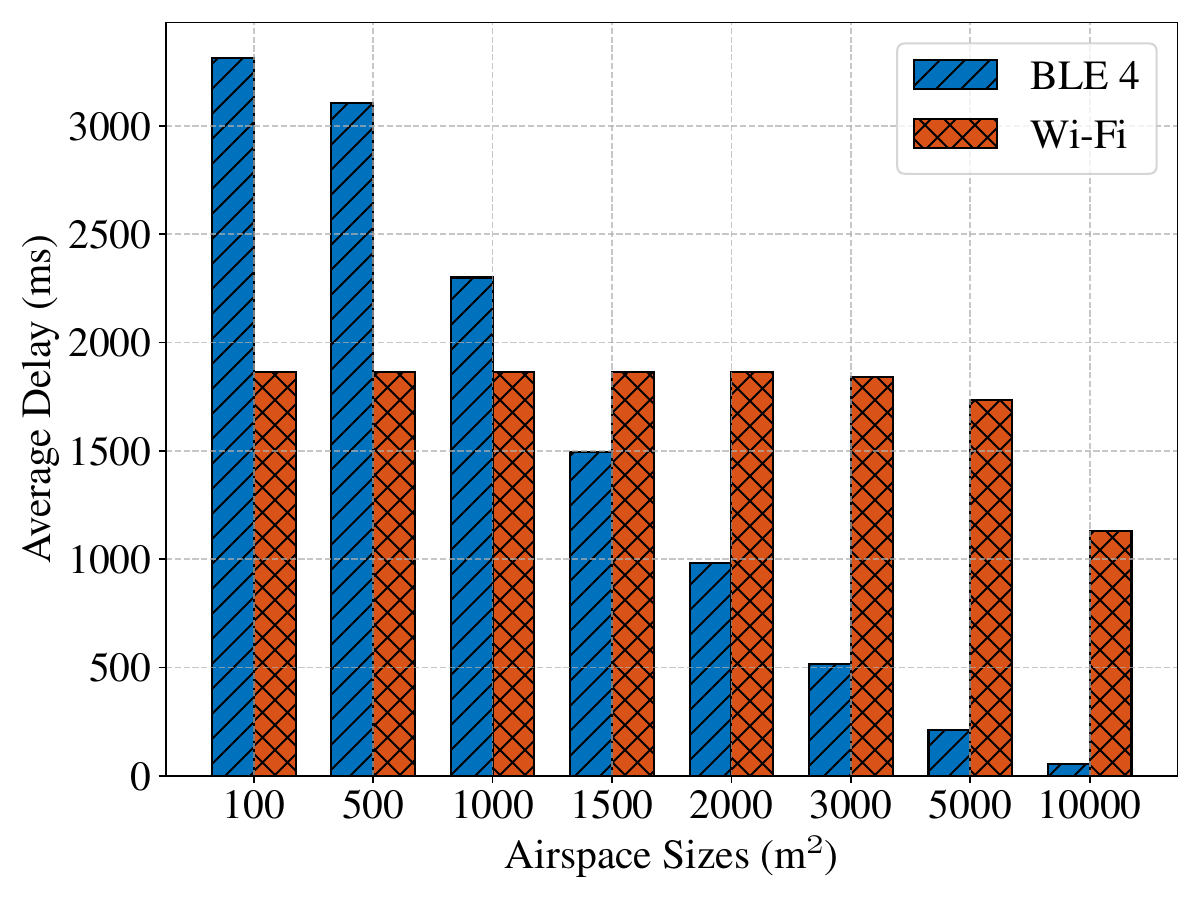}}
  \caption{Analysis of fixed Remote ID modes. (a) Average system transmission delay for BLE 4 and Wi-Fi across varying
  airspace sizes and message transmission rates. (b) Average system transmission delay for BLE 4 and Wi-Fi under optimal
  message transmission rate across different airspace sizes. \label{fig:fixed}}\vspace{-3mm}
\end{figure}

Fig. \ref{fig:fixed}(a) shows how the transmission delay varies for BLE 4 and Wi-Fi as the message transmission rate changes across different airspace sizes. Generally, higher transmission rates reduce delays by improving reception success during each GNSS update cycle. However, both protocols experience an increase in transmission delay at certain rates due to mismatches between transmission and reception cycles \cite{BLE4.0}. With current configurations, BLE 4 performs optimally at a rate of 9 messages/s, while Wi-Fi performs best at a rate of 10 messages/s.

Fig. \ref{fig:fixed}(b) presents the average system transmission delay for BLE 4 and Wi-Fi with their optimal rate settings. 
We specifically select several representative airspace sizes to illustrate the impact of different protocols on the average transmission delay of UAVs. In relatively high-density environments, such as those with airspace horizontal sizes of 100m$^2$, 500m$^2$, and 1,000m$^2$, Wi-Fi achieves lower delays than BLE 4. It indicates that in more congested airspace scenarios, Wi-Fi may be a more favorable choice for reducing communication delay.
Conversely, in relatively low-density environments, for example, with airspace sizes of 1,500m$^2$, 2,000m$^2$, and 3,000m$^2$, BLE 4 performs better in terms of average transmission delay. As the airspace density further decreases in larger airspace sizes like 5,000m$^2$ and 10,000m$^2$, the advantage of BLE 4 becomes even more evident. These results  highlight the potential benefits of adaptive mode switching for optimizing communication performance.

\begin{figure*}[t]
  \centering
  \subfloat[]{\includegraphics[width=0.29\textwidth]{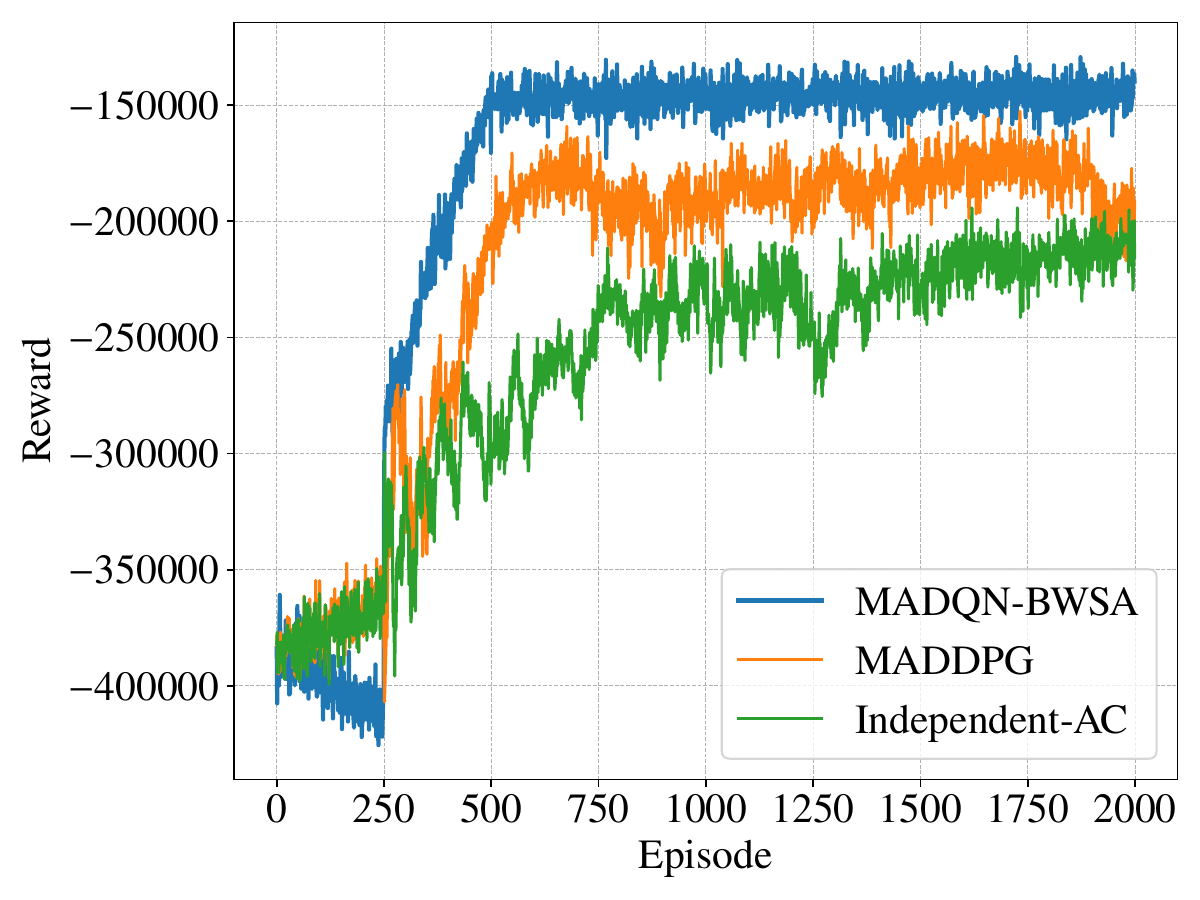}}%
  \subfloat[]{\includegraphics[width=0.29\textwidth]{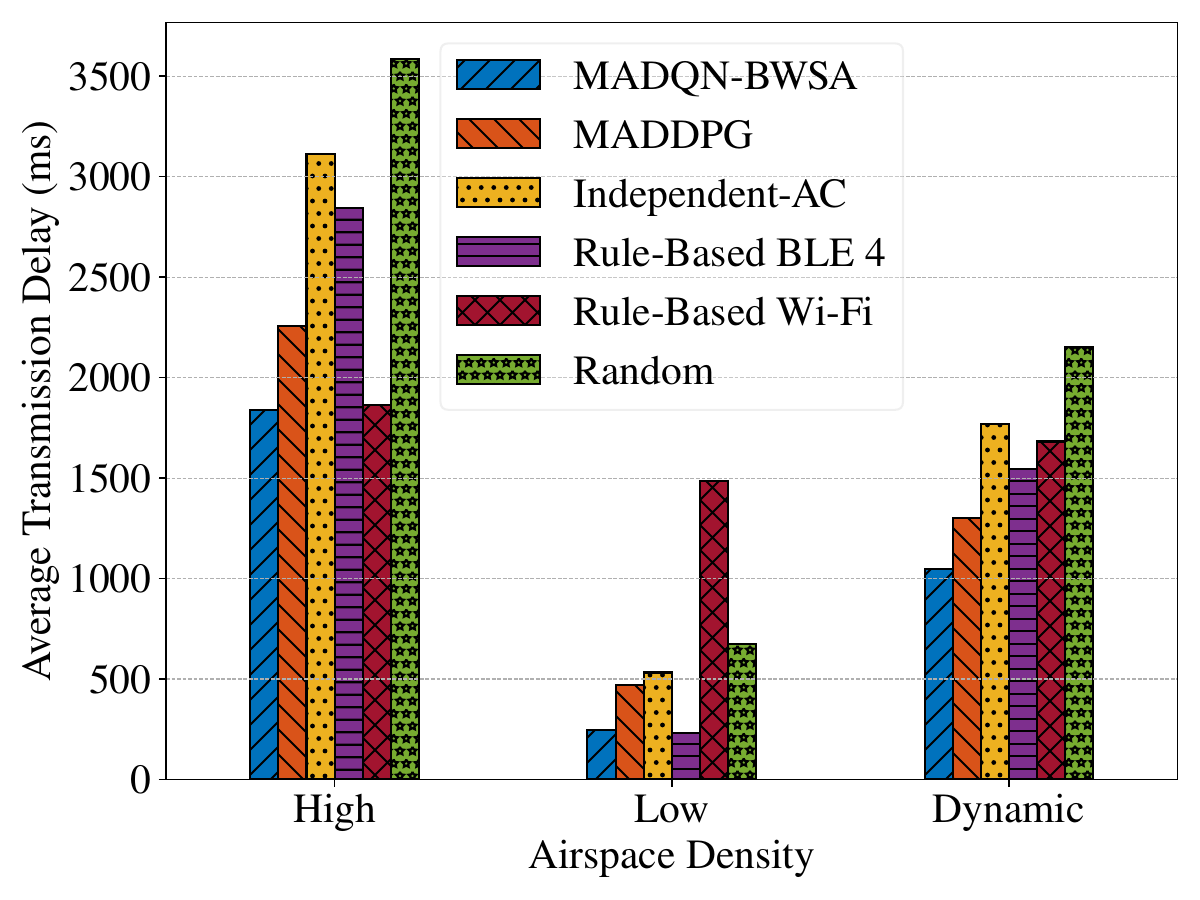}}
  \subfloat[]{\includegraphics[width=0.28\textwidth]{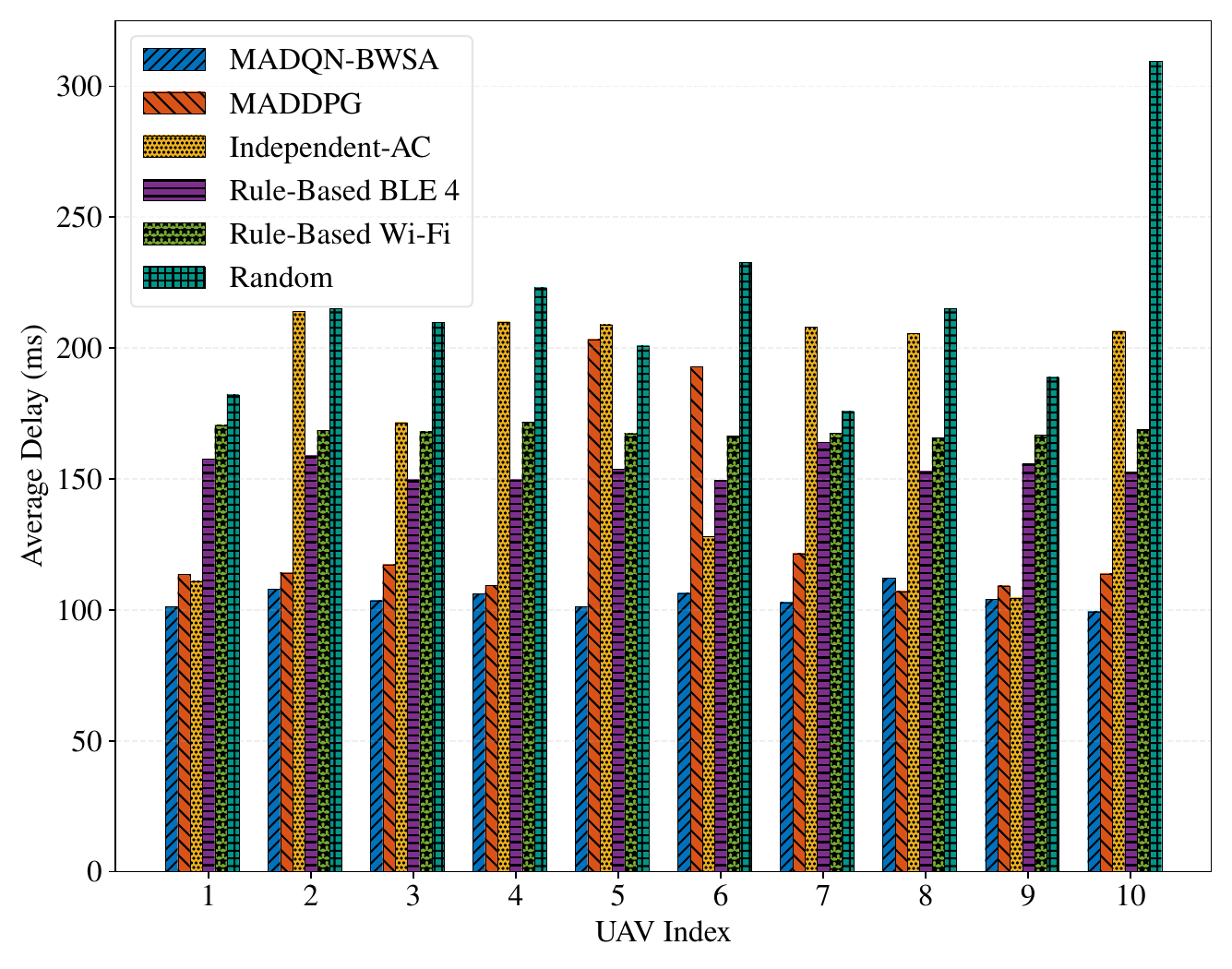}}%
  \caption{Analysis of MADQN-BWSA. (a) Reward convergence of MADQN-BWSA and baselines. (b) Average system transmission delay of different methods across various airspace densities. (c) Average transmission delays of each UAV in dynamic airspace density
  across different methods.\label{fig:algo}}\vspace{-3mm}
\end{figure*}
\vspace{-2mm}
\subsection{Analysis of MADQN-BWSA}
We evaluate the performance of the MADQN-BWSA in optimizing transmission delay. The UAV altitude is restricted between 30m and 120m, and horizontal airspace varies from 100m$^2$ to 10,000m$^2$. Simulations are conducted using Python 3.9 and TensorFlow 2.10. Training parameters include a batch size of 256, a replay buffer size of 25,000, and a discount factor of 0.95. The Adam optimizer is used with a learning rate of 0.0001, a soft update rate of 0.999, and an initial exploration rate of 1, decaying to 0.1 over 500 episodes. The total number of episodes is 1,000, with 100 steps per episode. The network has two hidden layers, with 256 neurons for the first layer and 128 for the second layer, and the weights for local and global rewards are both set to 1.

To assess the algorithm, we compare it with two baselines: the multi-agent deep deterministic policy gradient (MADDPG), which uses centralized training with decentralized execution and incorporates global action information, and the independent actor-critic (Independent-AC), which follows a policy-based approach with both distributed training and execution.

Fig. \ref{fig:algo}(a) shows the performance and convergence of the three algorithms. While all achieve convergence, MADQN-BWSA outperforms the others in both the speed and reward stability, with the baseline algorithms showing fluctuations after convergence.

Fig. \ref{fig:algo}(b) illustrates the system-wide average transmission delay for 10 UAVs across different airspace horizontal ranges. "High airspace density" includes UAVs in 100m$^2$, 500m$^2$, and 1,000m$^2$ airspaces, while "low airspace density" includes 3,000m$^2$, 5,000m$^2$, and 10,000m$^2$ airspaces. "Dynamic airspace density" involves both high-density and low-density airspace environments. Results are compared to fixed transmission configurations with the optimal message transmission rate for each protocol.
In high-density scenarios, our algorithm achieves a system-wide average delay of 1838ms, compared to 1865.1ms for rule-based Wi-Fi, demonstrating the ability of UAVs to autonomously switch to Wi-Fi mode and optimize the transmission rate. In low-density scenarios, our algorithm further reduces the average delay to 245.5ms, whereas rule-based BLE 4 achieves 232.3ms, indicating the UAV switches to BLE 4 transmission mode.
In dynamic-density scenarios, the advantage of mode switching is particularly evident. Our algorithm achieves a system-wide average delay of 1045.3ms, significantly lower than the fixed transmission modes of BLE 4 (1,544.7ms) and Wi-Fi (1,681.8ms), with reductions of 32.1\% and 37.7\%, respectively.
Fig. \ref{fig:algo}(c) shows the average transmission delay of each UAV in the dynamic-density environment. It can be seen that the proposed algorithm maintains an average and stable result among the baseline algorithms.\vspace{-2mm}

\section{Conclusions\label{sec:Conclusions}}
This paper focuses on optimizing the trnamission delays in the Remote ID enabled UAV communications. An average message transmission delay model is developed for the BLE 4 and Wi-Fi broadcast protocols, considering packet reception delays and collisions. Then, the MADQN-BWSA for dynamic communication protocol switching is proposed. Simulation results demonstrate that the proposed algorithm reduces UAV Remote ID communication delays by 32.1\% and 37.7\%, respectively, compared to fixed BLE 4 and Wi-Fi transmission modes in dynamic environments. 
\vspace{-2mm}
\textcolor{black}{\bibliographystyle{IEEEtran}
\bibliography{ref}}
\end{document}